\documentclass[a4paper,11pt,fleqn]{article}

\usepackage[ansinew]{inputenc}
\usepackage{hyperref}
\usepackage[mathscr]{eucal}
\usepackage{graphicx}
\usepackage[font=footnotesize,labelsep=period,justification=justified,figureposition=bottom,tableposition=top]{caption}
\usepackage{subcaption}
\usepackage{multirow}
\usepackage{amsmath,amssymb,amsthm}
\usepackage{comment}
\usepackage{ bbold }
\usepackage{hyperref}
\allowdisplaybreaks

\hypersetup{colorlinks, linkcolor=blue, citecolor=blue, urlcolor=blue}
\newcommand{\todo}[1][\null]{\ensuremath{\clubsuit}}

\newcommand{\noprint}[1]{}

{\theoremstyle{definition}

\newtheorem*{remark*}{Remark}
}

\newcommand{\checked}[1][\null]{\ensuremath{\boldsymbol{\surd}}}

\begin{document}

\par\noindent {\LARGE\bf
Towards replacing precipitation ensemble predictions systems using machine learning 
\par}

\vspace{4mm}\par\noindent {\large
R\"udiger Brecht$^{\dag}$ and Alex Bihlo$^\ddag$
\par}

\vspace{4mm}\par\noindent{\it
$^{\dag}$Department of Mathematics, University of Hamburg, Hamburg, Germany
}

\vspace{2mm}\par\noindent{\it
$^{\ddag}$ Department of Mathematics and Statistics, Memorial University of Newfoundland,\\
$\phantom{^{\ddag}}$~St.\ John's (NL) A1C 5S7, Canada
}

\vspace{2mm}\par\noindent {\it
\textup{E-mail:} ruediger.brecht@uni-hamburg.de, abihlo@mun.ca
}\par

\vspace{12mm}\par\noindent\hspace*{10mm}\parbox{140mm}{\small
Precipitation forecasts are less accurate compared to other meteorological fields because several key processes affecting precipitation distribution and intensity occur below the resolved scale of global weather prediction models. This requires to use higher resolution simulations. To generate an uncertainty prediction associated with the forecast, ensembles of simulations are run simultaneously. However, the computational cost is a limiting factor here. Thus, instead of generating an ensemble system from simulations there is a trend of using neural networks. Unfortunately the data for high resolution ensemble runs is not available. We propose a new approach to generating ensemble weather predictions for high-resolution precipitation without requiring high-resolution training data. The method uses generative adversarial networks to learn the complex patterns of precipitation and produce diverse and realistic precipitation fields, allowing to generate realistic precipitation ensemble members using only the available control forecast. We demonstrate the feasibility of generating realistic precipitation ensemble members on unseen higher resolutions.
We use evaluation metrics such as RMSE, CRPS, rank histogram and ROC curves to demonstrate that our generated ensemble is almost identical to the ECMWF IFS ensemble.
\par}\vspace{7mm}

Key points:
\begin{itemize}\itemsep=0ex
    \item  We propose a novel method for generating precipitation ensemble members from  deterministic weather forecasts.
    \item Prediction is done using a generative adversarial network in an image-to-image style.
    \item The neural networks is trained on low resolution data but can be applied on unseen higher resolution data.
\end{itemize}

\section{Introduction}
Precipitation forecasting is an essential aspect of weather prediction, with significant implications for various sectors, such as agriculture, transportation, water management, and disaster preparedness \cite{roebber1996complex,yu2016impact}.
In recent decades, the quality of numerical weather prediction has significantly improved, with meteorological centers utilizing numerical models and reanalyses with grid spacing ranging from 10 to 80 km and updated in near real-time \cite{feser2011regional}. However, while these grid spacing are capable of resolving large-scale weather patterns, they are insufficient for accurately representing precipitation in regions with  subgrid-scale orographic variations \cite{holden2011empirical}.
Thus, precipitation intensity can exhibit significant variations over short distances, with scales of 1 km or less, which is far finer than the typical resolution of global weather models. It is imperative to enhance the resolution of precipitation forecasts to accurately evaluate potential impacts, particularly for scenarios involving extreme rainfall.

Moreover, ensemble weather prediction aims to quantify the different sources of uncertainty in numerical weather prediction models, cf.~\cite{leutbecher2008ensemble}. The most significant sources of uncertainty are the initial conditions and errors in the numerical model formulation. To address these uncertainties, an ensemble of perturbed forecasts is generated, in addition to the single deterministic weather forecast, with the overall divergence, or spread, of the ensemble ideally providing a measure of the uncertainty in the deterministic prediction. However, the main constraint in generating the ensemble is still computational, as each ensemble run requires significant computational resources, thereby limiting the total number of ensembles that can be computed on an operational basis, to typically less than 100.

This article presents a generative deep learning approach to generating ensemble weather forecasts for high-resolution precipitation forecasts, without being trained on high-resolution data.
We use generative adversarial networks (GANs) to learn the complex spatio-temporal patterns of precipitation and generate diverse and realistic precipitation fields. 

Thus, the main contributions of this paper are:
\begin{itemize}\itemsep=0ex
    \item Using machine learning to generate realistic precipitation ensemble members from using only the available control forecast;
    \item Showing that it is possible to generating realistic precipitation ensemble members on unseen (higher) resolutions using lower resolution training data.
\end{itemize}

The remainder of this paper is organized as follows. In Section \ref{sec:relatedwork}, we provide a brief overview of related work on deep learning in meteorology and ensemble forecasting for precipitation. Section~\ref{sec:method} is devoted to the description of our proposed method using a generative deep learning approach. In Section \ref{sec:results} we present the results and verification of our approach. Finally, in Section~\ref{sec:conclusion}, we present a summary of the work carried out and discuss future directions for research in this area.

\section{Related work}\label{sec:relatedwork}

With the advent of modern deep learning following the seminal contribution~\cite{kriz12a}, neural networks have become a topic of immense research interest in meteorology. Neural networks have been used for precipitation nowcasting~\cite{xing15a}, data-driven weather forecasting~\cite{bihlo2021generative, path22a, weyn19a}, parameterization~\cite{gentine2018could}, as new numerical methods~\cite{bihlo2022physics,brech23b}, to improve Lagrangian particle dispersion models~\cite{brecht2022improving}, and data assimilation~\cite{geer21a}. Deep learning based models have also been used to explore a range of data-driven ensemble prediction methods. In~\cite{bihlo2021generative}, a conditional generative adversarial network was used to predict various meteorological quantities, with the ensemble prediction model being generated using Monte-Carlo dropout. 
In~\cite{scher2021ensemble}, several methods for generating ensemble predictions using deep learning were investigated, which included random perturbations, retraining of neural networks, Monte-Carlo dropout, and ensembles based on singular value decomposition. The work~\cite{gronquist2019predicting} explored the use of deep convolutional neural networks to reduce the number of ensemble members while preserving the correct ensemble spread, while \cite{gronquist2021deep} proposed post-processing computed ensemble members using neural networks. Ensemble forecasting based on convolutional neural networks on a cubed sphere for sub-seasonal forecasting was suggested in~\cite{weyn2021sub}. The ensemble generation by stacking sub-neural networks trained on different subsets of predictive meteorological variables was put forth in~\cite{clare2021combining}. A framework based on physically constrained generative adversarial networks to simultaneously improve local distributions and spatial structures of ensemble predictions was proposed in~\cite{hess2022physically}. The possibility of directly generating ensemble weather forecasts from the deterministic weather prediction only was shown in~\cite{brech23a}, who successfully used it to obtain ensemble weather predictions for the 500 hPa geopotential height. Crucially, this approach only generated the ensemble spread, not individual ensemble members. 
It is this approach that will be extended here to a full precipitation ensemble prediction system, whereby the control forecast is used to generated different ensemble members, rather than just the spread derived from these members.

Finally, also closely related to our work here, in \cite{harris2022generative} different generative neural network architectures are compared to stochastically down-scale a precipitation forecast. This way a high resolution ensemble forecast can be generated. However, this method relies on high resolution radar data samples, which are not available for all spatial locations where ensemble forecast would be needed.

\section{Methods}\label{sec:method}

In this section we describe the neural network architectures and the data being used to train these networks. 

\subsection{Data} \label{sec:data}
 
 We use the operational ECMWF IFS precipitation data. We downloaded the data on a regular longitude--latitude grid over northern Europe, with latitude $\lambda\in[45,73.5]$ and longitude $\varphi\in[-27,33]$ with a grid resolution of $288\times 513$ grid points. We retrieved the ensemble data initialized at 00 UTC, for the next 72 hours at a temporal resolution of 6 hours corresponding to 12 time steps per ensemble run, from the years 2020--2022. We use the years 2020 and 2021 for training and 2022 for testing. For training we randomly split the data into 95$\%$ for training and $5\%$ for validation.

 We note here that for deep neural networks is a rather modest training dataset. However, based on the previous work in~\cite{bihlo2021generative,brech23a,harris2022generative}, where training datasets of comparative size were used, we also found for the present study that this already allowed us to train highly performing neural network models.

\subsection{Neural network architecture} 

Our method is such that each model accepts as input the deterministic precipitation forecast for the next 72 hours, and predicts a a single ensemble member for the same time period. We then train a total of 10 such models, which collectively form our ensemble prediction system. We have chosen to generate 10 ensemble members rather than 50 members, as is done operationally at the ECMWF, since this provided a good compromise between training effort and statistical variability of the ensemble.

Alternatively, one could train a single model with a total of 10 output heads, but we found experimentally that the former method gives rise to a more diverse ensemble. While our method incurs a substantially higher training time, training only has to be done once and can be trivially parallelized. Our method also allows us to add further ensemble members without having to retrain the models for the existing ensemble members.

We interpret the precipitation forecast as a set of 12 image snapshots and note that the precipitation at a given time is the cumulative precipitation until that time. We consider a model that is based on the classical \texttt{pix2pix} architecture adapted from \cite{isola2017image}, where it was used for a variety of image-to-image translation problems. This model was selected as it has successfully been used in the related works~\cite{bihlo2021generative, brech23a}. It constitutes a conditional generative adversarial network, which consists of a generator and a discriminator network, which are trained in tandem through a two-player minimax game, see~\cite{good14a} for more information. 

\begin{figure}
    \centering
    \includegraphics[width=0.7\textwidth]{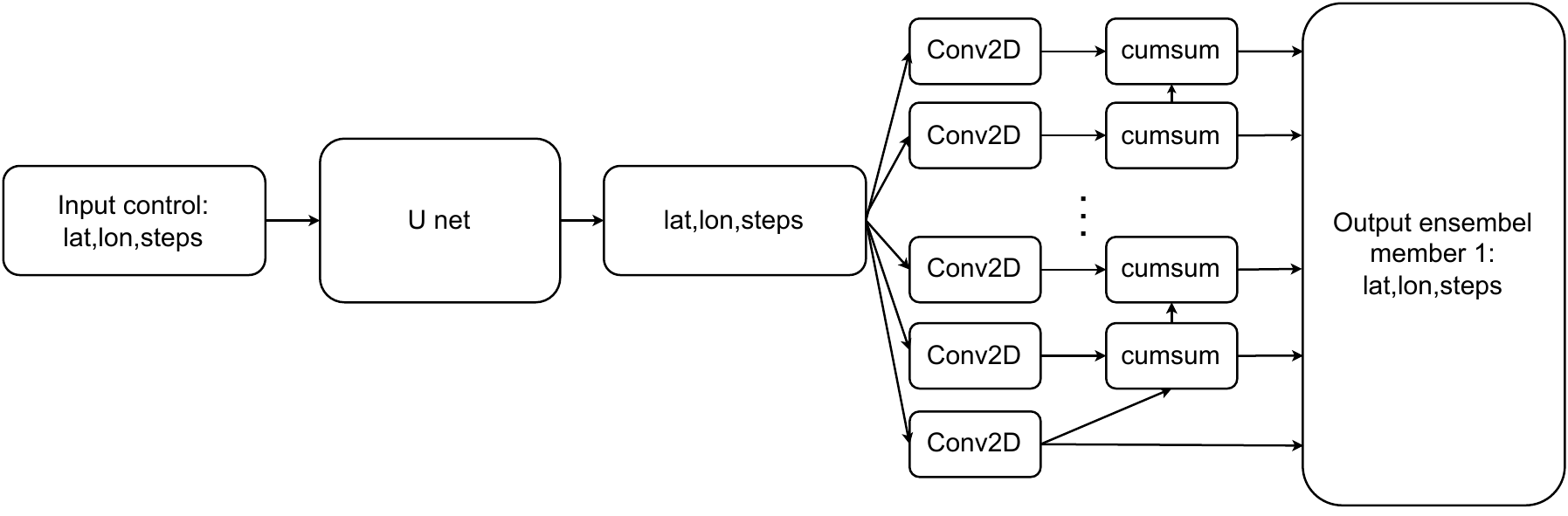}
    \caption{The neural network architecture of the generator network. For the architecture of the U-net, see Table~\ref{tab:generator}. The output of the U-net is fed into 12 parallel convolutions, one for each time step, with a single filter of size 3 and a ReLU activation. To guarantee that the cumulative rain is increasing over each time step, the output at each time step is formed by adding the respective convolutional output to the previous cumulative output.}  
    \label{fig:architecture}
\end{figure}

The neural network generator architecture is visualized in Fig.\ref{fig:architecture}. The discriminator architecture is the same as those used in~\cite{bihlo2021generative, brech23a}, with 2D convolutions instead of 3D convolutions and is thus not shown here. The details\footnote{The code can be found on \href{https://github.com/RudigerBrecht/Towards-replacing-precipitation-ensemble-predictions-systems-using-machine-learning-}{Github} upon request.} of the architectures of both the generator and discriminator can be found in tables~\ref{tab:generator} and~\ref{tab:discriminator}.

We train for each of the 10 ensemble members a neural network on both low resolution input (control forecast) and output (ensemble) data to predict the respective ensemble member given only the control forecast as input. The evaluation of the trained network is then done on the full high resolution input and output data.

Our reasoning behind using degraded data is to mimic the unavailability of high-resolution ensemble runs. We also note that as operational forecasting centers move towards higher resolution models, the respective ensemble predictions become computationally more expensive as well. If an ensemble prediction system pre-trained on lower resolution data still performs well on higher resolution data, this can help ease the computational and environmental burden of such systems.

Practically, since we use a convolutional neural network for the generator, which does not require a specification of the spatial input data dimensions, the trained neural network can be applied to arbitrary resolution data. The only constraint here is that the input and output data have to be integer multiples of the original training data. In the following, we will experiment with training our model on data that uses only every second and every third grid point, respectively, while being evaluated on the full high resolution testing dataset. 

\begin{table}[!ht]
\caption{Implementation details of the U-net neural network. The first three deconvolution layers are concatinated with the output of the convolution layers to form the skip connections.}
\begin{center}
\begin{tabular}{l||c|c|c|c}
U-net            & no.\ of filter & kernel size & strides & activation \\ \hline \hline
2D convolution & 64     & 4       & 2 & LeakyReLU \\
batch normalization & & & & \\
2D convolution & 128     & 4       & 2 & LeakyReLU \\
batch normalization & & & & \\
2D convolution & 256     & 4       & 2 & LeakyReLU \\
batch normalization & & & & \\
\hline
2D deconvolution & 256     & 4       & 2 & ReLU \\
batch normalization & & & & \\
2D deconvolution, 50\% dropout & 128     & 4       & 2 & ReLU \\
batch normalization & & & & \\
2D deconvolution, 50\% dropout & 64     & 4       & 2 & ReLU \\
batch normalization & & & & \\
2D deconvolution & 12     & 4       & 2 & ReLU \\
\end{tabular}
\end{center}
\label{tab:generator}
\end{table}

\begin{table}[!ht]
\caption{Implementation details of the discriminator neural network.}
\begin{center}
\begin{tabular}{l||c|c|c|c}
Discriminator            & no.\ of filter & kernel size & strides & activation  \\ \hline \hline
2D convolution & 64     & 4    & 2 & LeakyReLU\\
2D convolution & 128    & 4    & 2 & LeakyReLU\\
2D convolution & 256    & 4    & 2 & LeakyReLU\\
2D convolution & 512    & 4    & 1 & LeakyReLU\\
batch normalization & & & & \\
2D convolution & 1    & 4    & 1 &  Linear \\
\end{tabular}
\end{center}
\label{tab:discriminator}
\end{table}
\subsection{Training}

We train three sets of neural networks, in the following referred to as \textit{nn1}, \textit{nn2} and \textit{nn3}, respectively, one for the full resolution, one for degraded data where we take every second grid point and one where we take every third grid point. We implement the models in TensorFlow 2.8
and trained them simultaneously on a machine with a dual NVIDIA RTX 8000 GPU. We use the Adam optimizer with a learning rate of $2\cdot10^{-4}$ using a batch size of 1. We trained the models for 30 epochs using early stopping, saving the model after each epoch, and chose the one with the lowest validation error. One training epoch for all models took about 20 minutes.

\subsection{Ensemble verification methods}

We evaluate how well our model can reproduce the IFS ensemble by evaluating how similar the respective ensemble distributions are. Specifically, we choose as error measures:
\begin{itemize}\itemsep=0ex
    \item The root mean squared error (RMSE),
    \item The continuous ranked probability score (CRPS) \cite{hersbach2000decomposition},
    \item The Receiver Operating Characteristic (ROC) curves, and
    \item The rank histogram.
\end{itemize}
   Below we give the definition of the RMSE and the CRPS,
\begin{align*}
	\text{RMSE}(x,y) &=\sqrt{\overline{(x-y)^2}},
	\\
	\text{CRPS}(F,y) &= \int_{-\infty}^\infty \left(F(x)-\mathbb{1}(x-y)\right)^2\mathrm{d}x.	
\end{align*}
Here, the overbar denotes spatial averaging and $F$ is the cumulative distribution function of the forecast density function, and $\mathbb{1}$ is the Heavyside step function. 
The smaller the RMSE and CRPS the better the result is.

The ROC curve evaluates the effectiveness of a binary classifier by plotting the true positive rate (sensitivity) against the false positive rate (1 $-$ specificity) over a range of probability thresholds~\cite{wilk11a}. In order to create ROC curves for a specific precipitation intensity, we generate an ensemble of predictions. For each pixel, we determine the percentage of ensemble members that predicted rainfall above the specified intensity, which we interpret as the probability of the event occurring. Each point on the ROC curve represents the performance of the system at a particular probability threshold used to differentiate positive from negative predictions. The ROC curve summarizes the system's performance across all probability thresholds, ranging from 0 to 1.

The rank histogram \cite{hamill2001interpretation} is generated by repeatedly tallying the rank of the verification (usually an observation) relative to values from an ensemble sorted from lowest to highest. A well-calibrated ensemble results in an uniform histogram. Here, however, we are only interested in verifying that the generated ensemble has a similar histogram compared to the IFS forecast ensemble, as this is the data our models were trained on. Improving the bias in the ensemble prediction system itself is beyond the scope of this work.

\section{Results}\label{sec:results}

In this section we demonstrate that it is possible to learn the ensemble members on a low resolution, enabling to predict the associated ensemble members on a higher resolution. For this we evaluate the trained neural networks on the unseen high-resolution data of 2022.

\subsection{RMSE}

\begin{figure}
    \centering
    \includegraphics[width=\textwidth]{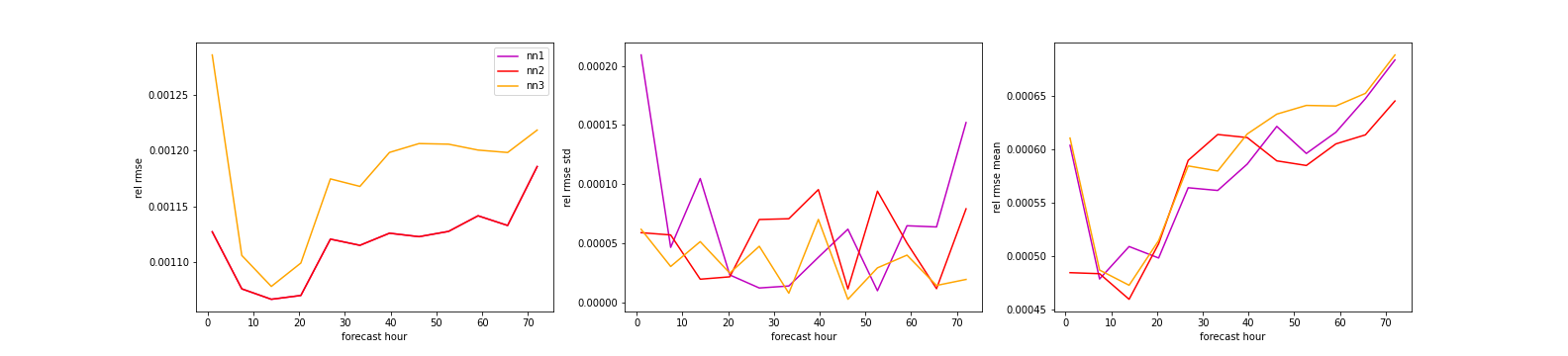}
    \caption{Forecast verification RMSE over the test dataset as a function of the forecast step.}
    \label{fig:rmse}
\end{figure}

 We compute the RMSE between the IFS ensemble and the ensemble generated for each of the 12 lead times and average the RMSE over the year. We also compute the RMSE between the ensemble mean and standard deviation.
In Fig. \ref{fig:rmse} we can observe that the error between the IFS ensemble and generated ensemble does not differ much between the low resolution (where it is trained on) and the higher resolution. 

\subsection{CRPS}
\begin{figure}
    \centering
    \includegraphics[width=0.6\textwidth]{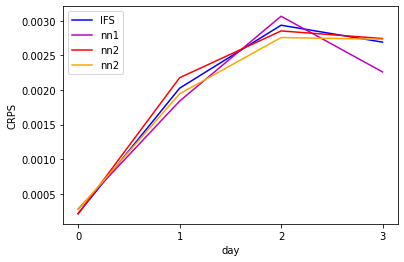}
    \caption{CRPS scores for the generated and IFS ensembles after 0, 1, 2 and 3 days
with the control run step 0 at those days as a reference.}
    \label{fig:crps}
\end{figure}

In Fig.~\ref{fig:crps} we show the CRPS value after 0, 1, 2 and 3 days averaged over the testing dataset. Here, we take the first step of the control run started after 0, 1, 2 and 3 days as a reference and compute the precipitation at those times based on the IFS ensemble and generated ensemble. We see in Fig.~\ref{fig:crps} that the CRPS values are almost the same, also for the different resolutions. 

\subsection{ROC curve}
\begin{figure}
    \centering
    \includegraphics[width=0.49\textwidth]{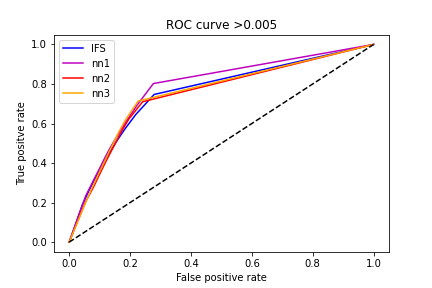}
    \includegraphics[width=0.49\textwidth]{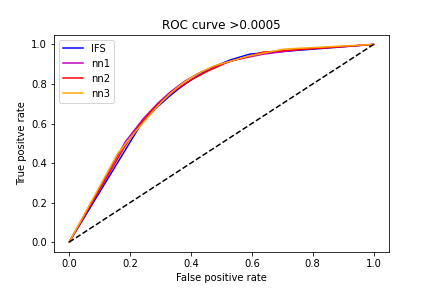}
    \caption{The ROC curves for the IFS and generated ensembles. The curves are computed after 2 days. Left for 0.005 and right for 0.0005 precipitation thresholds. The dashed diagonal line represents random chance, such that only points above it are a meaningful result. }
    \label{fig:roc}
\end{figure}

We use the data from the control run started after 2 days as the reference and compute the precipitation of the IFS and generated ensembles. We subtract the total precipitation at step 8 (day 2) of the step 4 (day 1). Then, we set all values of the reference below the thresholds (0.005 and 0.0005) to 0 and above to 1 to create a binary classification. We do the same for each of the 10 ensemble member to generate a probability for the specific rain rate. Then, we compute the mean of the ROC curves over the year 2022. In Fig. \ref{fig:roc} we see that the ROC curves of the generated ensembles agree with the one produced by the IFS ensemble.

\subsection{Rank histogram}

\begin{figure}
    \centering
    \includegraphics[width=\textwidth]{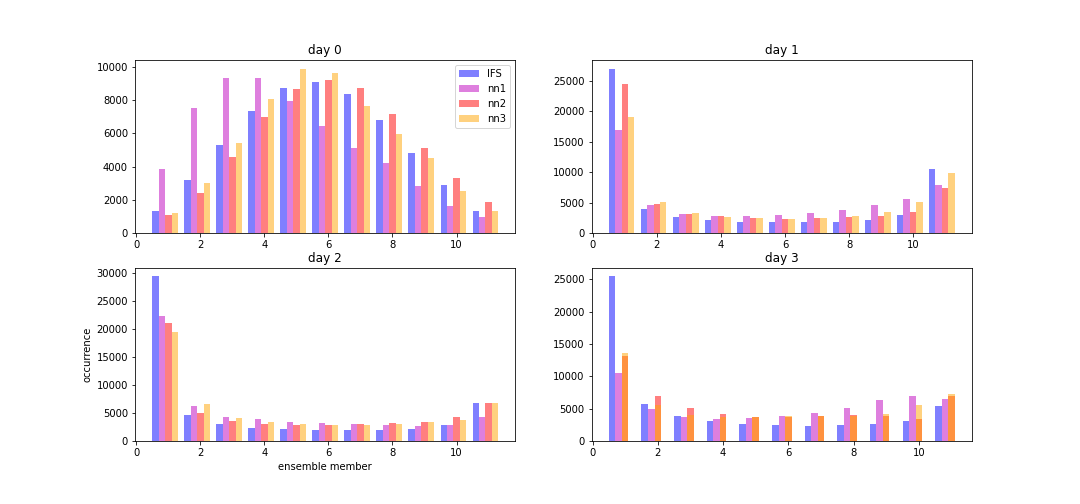}
    \caption{Comparison of the rank histograms for the generated and IFS ensemble after 0, 1,2 and 3 days
with the control run step 0 at those days as a reference.}
    \label{fig:ranky}
\end{figure}

With the rank histogram we want to evaluate if the generated ensemble has a similar calibrated ensemble spread compared to the IFS ensemble. 
We use the data from the first  control run step which is started after 0, 1, 2 and 3 days as the reference. Then we compute the difference with the previous day for the IFS and generated ensembles to get the same precipitation forecast. 
In Fig.~\ref{fig:ranky} we observe that the ensemble spread of the generated and IFS ensemble are very similar. The non uniform shape of the histogram could be related to using the control run started after day 0, 1, 2 and 3 as a reference. However, here we are only interested in the similarity of the ensemble.  

\subsection{Case study}

\begin{figure}
    \centering
    \includegraphics[width=\textwidth]{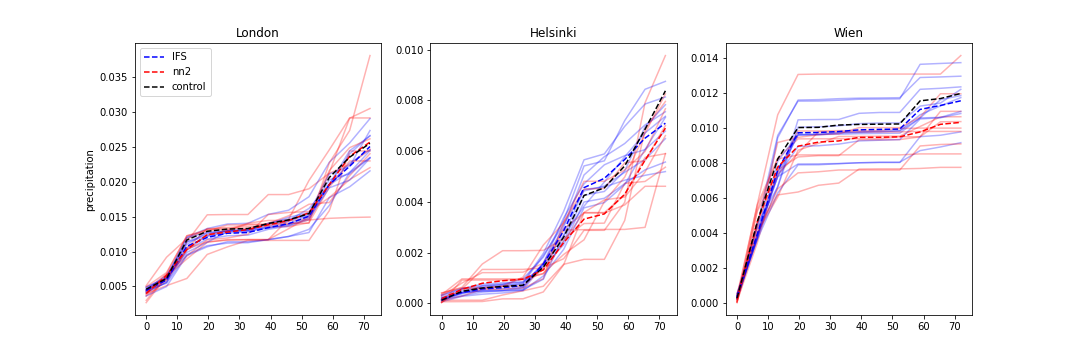}
    \includegraphics[width=\textwidth]{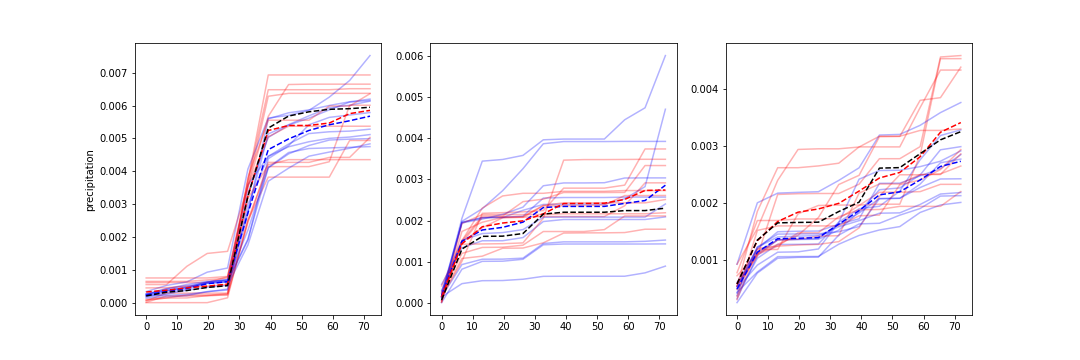}
    \includegraphics[width=\textwidth]{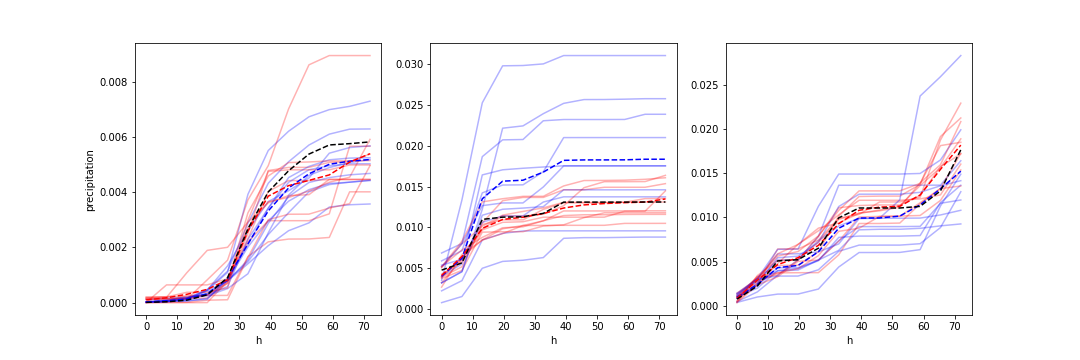}
    \caption{We show the IFS and generated ensemble members at different times of the year (January 15th: first row, April 25th: middle row, August 3rd: bottom row) in three different city. Here, we averaged the precipitation over a $10\times10$ pixel stencil around the city. The solid lines correspond to different ensmeble members and the dashed line to their mean.}
    \label{fig:case}
\end{figure}

\begin{figure}
    \centering
    \begin{subfigure}{\textwidth}
    \includegraphics[width=\textwidth]{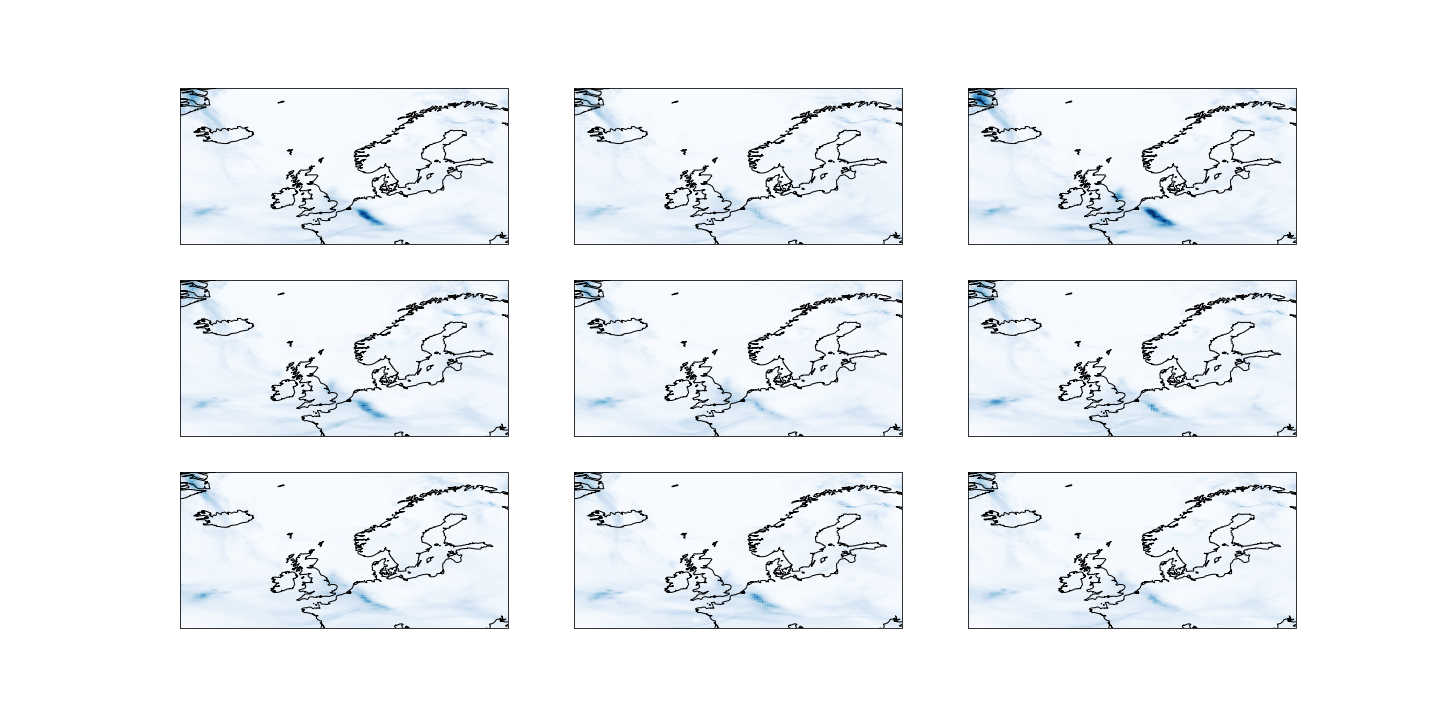}
    \caption{Generated ensemble members using the neural network \textit{nn2}.}
    \end{subfigure}
    \begin{subfigure}{\textwidth}
    \includegraphics[width=\textwidth]{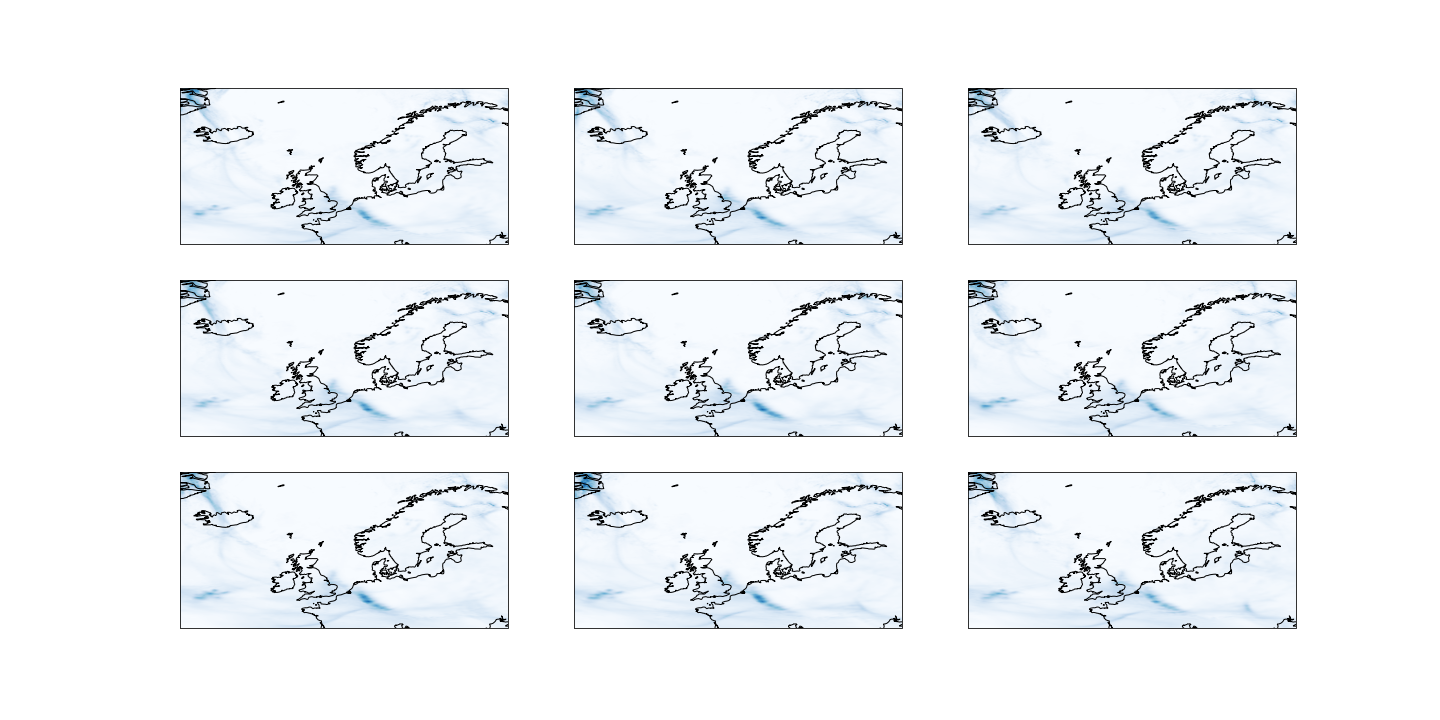}
    \caption{IFS ensemble members.}
    \end{subfigure}
    \caption{We show the IFS and generated ensemble members on 15th January using \textit{nn2}. The results for April 25th and August 3rd, and for \textit{nn3}, are both qualitatively and quantitatively similar and thus not shown here.}
    \label{fig:mapplot}
\end{figure}

To present a case study, we evaluate the precipitation ensemble members at different cities and compare them with the control run on three different randomly selected days, January 15th, April 25th and August 3rd. In Figs.~\ref{fig:case} and~\ref{fig:mapplot} we show the ensemble generated from the neural network \textit{nn2}, which was trained on the degraded data taking every second grid point. We note that the ensemble from the network \textit{nn1} and \textit{nn3} is qualitatively similar. In general the spread of the generated and IFS ensemble is very similar. 

\section{Conclusions}\label{sec:conclusion}

In this paper, we present a new method to generate ensemble members from one deterministic control run. We want to highlight that this method does not rely on any high resolution samples to generate ensemble members on higher resolution simulations. 
We trained a conditional generative adversarial network, an adaptation of the \texttt{pix2pix} model, for this purpose. Using different verification methods, we could demonstrate that the generated ensemble members have close statistical properties compared to the IFS ensemble.  

A limitation of this work is that precipitation events are globally sparse, such that it is difficult to train one accurate global model with this setup. We also note that precipitation has different characteristics in the tropics and extra-tropics, and data quality varies based on spatial locations~\cite{vill08a}. As a consequence, a new set of neural networks needs to be trained for each region of interest. Given the relatively short training times of the proposed neural networks and the continuous increase in dedicated artificial intelligence hardware, we do not believe though that this is a serious issue. Additionally, many practical applications only require accurate precipitations forecasts over limited spatio-temporal domains.

There are various possible extensions of this work. To improve the accuracy of predicting extreme precipitation events, utilizing more training data, and potentially more powerful neural network architectures, could be beneficial. In this paper we have used generative adversarial networks as the main architecture. It is well-known that these can suffer from training difficulties such as mode collapse leading to less diverse generated samples compared to architectures based on, e.g., diffusion models~\cite{li2022srdiff}. While we have not observed mode collapse in practice, and the generated ensembles have excellent statistical properties, for the further improvement of results, it may be beneficial to look into more advanced image-to-image translation architectures. 

We also anticipate that our model has the potential to be extended for broader application in different geographical regions, and for different meteorological parameters. It will also be interesting to see if the inclusion of additional input features, such as temperature, dew point, and wind speed, could further improve the obtainable results.

\section*{Acknowledgements}

The authors thank Thomas Haiden and Mariana Clare for helpful discussion. This research was undertaken thanks to funding from the Deutsche Forschungsgemeinschaft (DFG, German Research Foundation) - Project-ID 274762653 - TRR 181, the Canada Research Chairs program and the NSERC Discovery Grant program. The authors also acknowledge support from the ECMWF special project \textit{Mining 5th generation reanalysis data for changes in the global energy cycle and for estimation of forecast uncertainty growth with generative adversarial networks}.

{\footnotesize\setlength{\itemsep}{0ex}

}

\end{document}